# An Approach for Reviewing Security-Related Aspects in Agile Requirements Specifications of Web Applications

Hugo Villamizar, Amadeu Anderlin Neto, Marcos Kalinowski,
Alessandro Garcia
Software Engineering Laboratory
Pontifical Catholic University of Rio de Janeiro (PUC-Rio)
Rio de Janeiro, Brazil
{hvillamizar, aanderlin, kalinowski, afgarcia}@inf.puc-rio.br

Daniel Mendez
Software Engineering Research Lab Sweden
Blekinge Institute of Technology
Karlskrona, Sweden
daniel.mendez@bth.se

*Abstract*—Defects in requirements specifications can have severe consequences during the software development lifecycle. Some of them result in overall project failure due to incorrect or missing quality characteristics such as security. There are several concerns that make security difficult to deal with; for instance, (1) when stakeholders discuss general requirements in (review) meetings, they are often not aware that they should also discuss security-related topics, and (2) they typically do not have enough security expertise. These concerns become even more challenging in agile development contexts, where lightweight documentation is typically involved. The goal of this paper is to design and evaluate an approach to support reviewing security-related aspects in agile requirements specifications of web applications. The designed approach considers user stories and security specifications as input and relates those user stories to security properties via Natural Language Processing (NLP) techniques. Based on the related security properties, our approach then identifies high-level security requirements from the Open Web Application Security Project (OWASP) to be verified and generates a focused reading techniques to support reviewers in detecting defects. We evaluate our approach via two controlled experiment trials. We compare the effectiveness and efficiency of novice inspectors verifying security aspects in agile requirements using our reading technique against using the complete list of OWASP high-level security requirements. The (statistically significant) results indicate that using the reading technique has a positive impact (with very large effect size) on the performance of inspectors in terms of effectiveness and efficiency.

## I. INTRODUCTION

Requirements Engineering (RE) is an inherently complex part of software engineering. Misunderstandings and defects in requirements-related artifacts can easily lead to design flaws and cause severe and costly problems [24]. Agile requirements engineering relies on lightweight documentation and face-to-face collaborations between customers and developers [9]. Yet, agility does not necessarily compensate the problems of more plan-driven software process models. In fact, it can even make those problems more explicit if a key prerequisite for successful RE is not given: human-intensive exchange, collaboration, and trust [19]. In other words, agile RE has already helped to address some specific problems of RE, but it has also brought others to the surface [20].

Security is an essential non-functional requirement that requires special attention, inter alia, due to business needs to protect data. Much of security-related information is hosted on the internet, making web applications a target. Security requirements often appear throughout but also beyond the elicitation process so that stakeholders are often simply not aware of them. For instance, when stakeholders make decisions along the process meetings, they are often not aware that these decisions might also raise data protection-related issues [24]. This often leads to not specifying relevant security-related aspects.

However, the picture is even more challenging in agile methods. Several studies have identified problems that could result from the poorly detailed requirements specifications [9][34]. These problems can result in overall project failure due to incorrect or missing functionalities and/or quality characteristics. According to Eberlein and Leite [15], there is a need for agile methods to include techniques that make it possible to identify non-functional requirements early. There is also a need to describe them in such a way that their early analysis may happen, thus reducing the likelyhood of costly rework [30]. In that sense, agile RE should include more detailed requirements verification into the process [15][34].

Despite the apparent advantages of integrating security into agile contexts, a recent mapping study by Villamizar et al. identified that there is a lack of research on security requirements verification [42]. Such activities should be conducted to assure that agile requirements specifications are correct, consistent, unambiguous, and complete. This means properly covering security-related aspects, such as task-flow constraints, information flow, and assignment of administrative privileges.

Given this scenario, our work aims at contributing to closing the literature and industry gaps that exist with respect to security requirements verification in agile contexts. More specifically, our goal is to propose and evaluate an approach for reviewing security-related aspects in agile requirements specifications of web applications. To achieve this goal, we define the following Specific Goals (SG):

– SG1: We aim at developing an approach for reviewing security-related aspects in agile requirements specifications of web applications.
– SG2: We evaluate the approach to analyze whether it helps inspectors to review security-related aspects. To this end, we report on a controlled experiment to observe the impact on effectiveness and efficiency. Additionally, we also observe the perceived usefulness and ease of use.

As a result of our work, we found that using our approach had a positive impact (with very large effect size) when reviewing security-related aspects in agile specifications of web applications. Our results indicate significant differences when comparing the performance of inspectors using our approach versus other defect-based technique.

The remainder of this work is organized as follows. Section 2 introduces the background on agile requirements engineering and on how security verification is typically performed in this context. Section 3 introduces the approach we designed for reviewing security-related aspects in agile requirements specifications of web applications. Section 4 presents the study design used to evaluate the approach. The results are presented and discussed in Sections 5 and 6, respectively. Section 7 provides a discussion of limitations of our approach before we conclude with Section 8.

## II. BACKGROUND AND RELATED WORK

This section introduces the background on agile and security requirements and describes work related to security requirements verification in agile context.

### A. Agile Requirements

The term "agile requirements engineering" emerged in response to the agile manifesto in 2001. It is used to define the ''agile way'' of planning, executing and reasoning about requirements engineering activities [25]. Yet, not much is known about the challenges posed by the collaboration-oriented agile way of dealing with requirements engineering activities. Ramesh *et al.* [35] performed a multi-case study with 16 organizations that develop software using agile methods. They identified that agile RE practices resulted in challenges regarding neglected non-functional requirements, minimum documentation and no requirements verifications. The recent report from the "Naming the Pain in Requirements Engineering" (NaPiRE) initiative [20] extends the list of challenges with: (i) communication flaws between project teams and the customer, and (ii) underspecified requirements that remain too abstract and, thus, are not measurable. These observations give a picture on the difficulties of dealing with non-functional requirements in agile environments. It is reasonable to believe that security requirements are no different in this respect.

### B. Security Requirements (SR)

Software development should be conducted with security in mind at all stages and it should not be an afterthought [30]. Developing secure software is not a trivial task due to the lack of security expertise in development teams and the inadequacy of existing methodologies to support developers who are not security experts [24]. Yet, in the majority of software projects, security is often dealt with in retrospective when the system has already been designed and put into operation [31].

SR have traditionally been considered "quality" requirements [11], [13]. Like other quality requirements, they tend not to have simple yes/no satisfaction criteria (e.g., as acceptance criteria). Haley et al. [23] presents some challenges related to SR. First, people generally think about and express SR in terms of "bad things" (negative properties) to be prevented. It is very difficult, if not impossible, to measure negative properties. Second, for SR, the tolerance on "satisfied enough" is small, often zero, given the implications of non-compliance. Moreover, stakeholders tend to want security requirement satisfaction to be very close to yes. Third, the effort stakeholders might be willing to dedicate to satisfying SR also depends on the likelihood and impact of a failure to comply with them. This is even more challenging in the context of agile software projects because, apparently, functionality is the primary focus in agile processes while non-functional requirements are typically either ignored or ill-defined [34].

### C. SR Verification in Agile Context

In general, several authors have worked on quality assurance methods for verifying the quality properties of different artifacts. One of the most compared and evaluated methods, in several experiments and studies, are inspection techniques. In that direction, we can name reading support techniques for defect detection such as Perspective-Based Reading (PBR) [3], Defect-Based Reading (DBR) [21] and Use-Based Reading (UBR) [45] that are well known and established. As far as security inspections are concerned, little work has been published (e.g., [10][14][16]) on how to support inspectors with detailed reading support for reviewing security related aspects.

Elberzhager et al. [16] propose a model for security goals that involves guided checklists to support inspectors when checking security. They describe a step-by-step guide that results in questions to be checked by an inspector. This model is similar to our proposal because it works using a reading technique that supports the inspector on how to review security. However, there are differences. First, our approach focuses on verifying security requirements in early stages, i.e., right after requirements specification and within agile requirement artifacts. Second, our approach addresses high-level security requirements as defined by the Open Web Application Security Project (OWASP) [33], which provides a well-known industry standard on security. Furthermore, our proposal involves classifying the defects found by inspectors, providing a better understanding of the distribution of the problems.

Carver et al. [10] further describe a set of perspectives that provide security-specific questions for a requirements inspection. Two of them are part of the PBR technique (designer and tester). They also created a new perspective based on the needs of a black hat tester. In this additional perspective, the reviewer focuses on three types of security information: cryptography, authentication, and data validation. According to the authors, those types of information and the related questions were adapted for requirements from Araujo and Curphey's article on security code reviews [1]. However, due to the large number of software vulnerabilities and the variety of ways to deploy computer attacks, it is not enough to consider only three types of security controls. Indeed, the list is incomplete when compared to OWASP security properties and high-level security requirements.

In the agile context, the picture is even poorer. We are aware of only one study that it is part of the results of a recent systematic mapping [42]. Domah et al. [14] propose a lightweight methodology to address non-functional requirements (NFRs) early in agile software development processes. NFRs elicitation, reasoning, and validation are considered within that methodology. Regarding verification, it depends on a quantification taxonomy with different levels of decomposition for identifying quantified validation criteria for each NFR. However, this methodology does not offer specific guidance (e.g., questions, checklist) to support inspectors in identifying security related defects in requirements specifications. Hence, previous knowledge on security is required to take advantage of the methodology.

In summary, very few approaches exist to address the systematic analysis and detection of security problems, especially during early development stages, and the scenario is even worse in agile contexts.

### III. OUR APPROACH

In this section, we introduce our approach for reviewing security-related aspects in agile requirements specifications of web applications. The approach aims at addressing security in early stages of the agile software development lifecycle and considers requirements specified using user stories. The remainder of this section describes the assumptions, the scope delimitation, an overview of our approach, and a motivational example.

#### A. Assumptions

The approach was designed with some underlying assumptions in mind. These assumptions are as follows.

*Requirements are specified in a user story format*. Instead of a structured and fully described requirements document, requirements are typically written down via user stories involving free form or with some constraints [43]. For this reason, the approach is focused on the user story format. These stories are often expressed in a simple sentence using the role/feature/reason schema and structured as follows: *As a [role], I want to [feature], so that [reason]*.

*User stories are analyzed independently*. It makes sense to consider one user story at a time because, typically, in agile methods, the features that are part of the backlog are analyzed separately in order to receive a "definition of ready" in order to be included in a sprint [43].

*The OWASP represents a reliable baseline and standard of security guidelines*. OWASP has a strong focus on web applications, the target of our approach. OWASP concerns providing impartial, practical information about security in web applications to individuals, corporations, universities, government agencies, and other organizations worldwide. Many open source security-related tools (e.g., Sonar) and current research (e.g., [2]) on web application security use OWASP as a definitive reference. Hence, we consider the reliability of this project as a reasonable assumption.

#### B. Approach Scope Delimitation

Hereafter, we answer some potential questions to provide a further understanding on the intended approach.

*For whom is the approach intended?* Our approach was designed to support novice inspectors and security analysts. It provides them with a reading technique to assist the identification of defects related to security aspects in agile requirements specifications. According to Nerur [32], people with a high-level of competence are of vital importance in agile teams. Much of the knowledge in agile development is tacit and resides in the heads of the development team members [7]. Nevertheless, in an agile context, there is a focus on functionality. Competence in software security is typically not widely spread among agile practitioners [22]. Therefore, our approach focuses on providing a detailed reading technique to support novice inspectors. We believe that more experienced security analysts could still use the approach, but these are not in the scope of our evaluation.

*What security-related aspects does the approach cover?* We decided to focus on security properties and high-level security requirements as proposed by the OWASP [33]. These high-level requirements describe the most important security-specific features that architects and developers should include in every web application project [33]. The SQuaRE (the System and Software Quality Requirements and Evaluation) quality model [26] also define security-related sub-characteristics, which hereafter, for term compatibility, will also be referred to as security properties. We compared the security properties of OWASP and SQuaRE. OWASP contains three security properties: *confidentiality*, *integrity*, and *availability*. The SQuaRE quality model, on the other hand, contains five: *confidentiality*, *accountability*, *integrity*, *non repudiation*, and *identification and authentication*. Based on their definitions, all of the SQuaRE security properties can be mapped into the OWASP security properties.

For our final list of considered security properties, we used the OWASP properties with a single change, splitting confidentiality into two separate properties: (i) *confidentiality* and (ii) *identification and authentication*. *I.e.*, we separated identification and authentication from other confidentiality problems. Table 1 presents the security properties as considered by our approach.

TABLE 1. SECURITY RELATED PROPERTIES CONSIDERED BY OUR APPROACH.

| Security Property | Description |
|---|---|
| Confidentiality | Degree to which the data is disclosed only as intended. |
| Integrity | Degree to which a system or component prevents unauthorized access to, or modification of, computer programs or data. |
| Availability | Degree to which a system or component is operational and accessible when required for use. |
| Identification & Authentication | Degree to which the identity of a subject or resource can be proved to be the one claimed. |

*What types of requirements defects does the approach cover?* In RE, a defect can be defined as any problem of correctness and completeness with respect to the requirements, internal consistency, or other quality attributes [39]. A common defect taxonomy used when inspecting requirements [28] is the one proposed by Shull [37]. The defect types in this taxonomy are: omission, ambiguity, inconsistent information, incorrect fact, and extraneous information. However, we excluded the extraneous information defect type (which concerns specifying requirements that are not needed). This decision was taken because we use the OWASP high-level security requirements as a reference, while they are stated as mandatory for inclusion; they are not necessarily complete given that specific security needs may sprout for specific applications. Hence, given the impact that a missing security requirement can have on the application, we did not feel comfortable to recommend exclusions. Table 2 shows the types of defects covered by our approach and its definitions.

*What kind of review technique does the approach use?* Typically, developers and software analysts rely on *ad-hoc*

methods or checklists to analyze documents. In an *ad-hoc* (i.e. unstructured) reading process, the reader is not given directions on how to read. The result is that reviewers tend to build up skills in document understanding slowly based on individual experiences acquired over time [36]. For this reason, we decided to focus the review process of our approach on a reading technique to increase the effectiveness of individual reviewers by providing a systematic guide that can be used to examine, in our case, security-related aspects and consequently identify defects.

TABLE 2. DEFECT TYPE DEFINITIONS IN SCOPE OF OUR APPROACH.

| Defect Type | Definition |
|---|---|
| Omission (OM) | Necessary information about the system has been omitted from the requirements. |
| Ambiguity (AM) | A requirement has multiple interpretations due to multiple terms for the same characteristic, or multiple meanings of a term in a particular context. |
| Inconsistency (IS) | Two or more requirements are in conflict with one another. |
| Incorrect Fact (IF) | A requirement asserts a fact that cannot be true under the conditions specified for the system. |

*In which part of the lifecycle of agile methods can the approach be applied?* Agile methods are characterized by having iterative structures that should allow early delivery, continual improvement, and rapid and flexible response to change [5]. Hence, we envision that our approach is used just before a user story is defined as "ready" for codifying.

## C. Overview of our Approach

We propose our approach in two defined phases: (1) generating the focused reading techniques based on the agile requirements specification, and (2) following the reading technique to identify defects. These phases are illustrated in detail in Figure 1 and explained as follows.

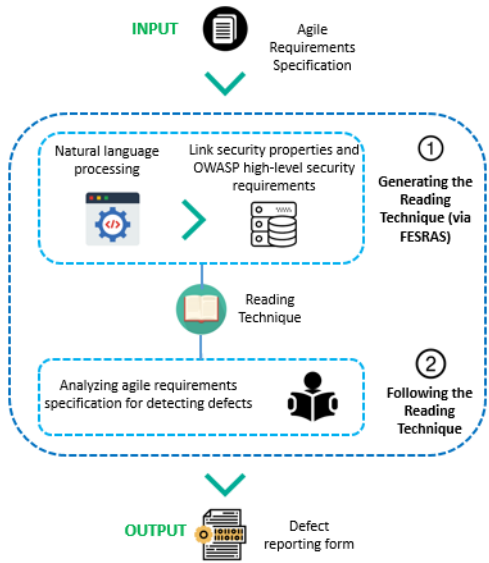

Fig. 1 Overall Structure of our Approach

*Phase 1: Generate Reading Technique.* In order to generate the focused reading technique, we use Natural Language Processing (NLP) to extract relevant words from the user story. Thereafter, these words are used to identify security properties and to link the related OWASP high-level security requirements to be verified. The availability of automatic tools for the quality analysis of Natural Language requirements is recognized as a key factor for achieving software quality [29]. Details on how relevant words and security properties are identified follow.

*Extracting Relevant Words.* This activity involves automatically analyzing user stories that describes features and functional requirements of the software to be built. Our approach extracts the relevant verb (action) of the user story that indicates a potential behavior be considered when thinking about security. In some cases, the nouns of the user story can also indicate situations where certain security features should be considered. Therefore, nouns are also extracted for matching purposes. The verb is extracted from the second block of the user story format and the nouns are extracted from the third block (see Table 3).

TABLE 3. WAY TO EXTRACT THE WORDS FROM THE USER STORY.

| Type of word | User story skeleton |
|---|---|
| Verbs | As a [user], I [**want to**], [so that]. |
| Nouns | As a [user], I [want to], [**so that**]. |

To extract the words, we developed a Software Framework for Eliciting Security Requirements from Agile Specifications (FESRAS)[1] that uses the Stanford CoreNLP tool[2] through a library. The Stanford CoreNLP provides a set of natural language analysis tools written in Java. It can take raw human language text input and provide the base forms of words. The library represents each sentence as a directed graph where the vertices are words and the edges are the relationships between them. Thereby, the software framework can take the verbs and nouns of the user story and then analyze them in order to link security properties.

*Identifying Security Properties and Linking High-Level Security Requirements.* After identifying the relevant words of the user story, we need to identify security properties in order to map high-level security requirements that represent a set of security-specific features to be verified. As an example, Table 4 shows some words that are part of the repository used to indicate which security properties should be considered. Our online material[3] contains all the words of the repository.

TABLE 4. SUBSET OF KEYWORDS AND ITS SECURITY PROPERTIES.

| Keyword | Security Property | | | |
| | Confidentiality | Integrity | Availability | Identification Authentication |
|---|---|---|---|---|
| Access | X | | | X |
| Change | | X | | |
| Export | X | | | |
| Send | X | | | |
| Recover | | | X | |
| Backup | | | X | |
| Password | | | | X |
| Role | | | | X |
| Time | | | X | |

This repository is strongly based on a similar one provided by Slankas and Williams [38] in their work about automated extraction of non-functional requirements in available documentation. The changes are that we focused on security and therefore we complemented the set of keywords with synonyms and more specific words stated by OWASP [33]. If there is no match between the extracted words and the security properties, our approach will automatically link the user story to all the security properties stored in the repository. In order to mitigate this scenario, synonyms of the initial set of keywords were added to the repository of keywords. Table 5 shows the OWASP high-level security requirements by security property.

---

[1] Framework URL temporarily omitted in order to comply with double-blind.
[2] https://github.com/stanfordnlp/CoreNLP
[3] https://doi.org/10.5281/zenodo.2602205

TABLE 5. OWASP HIGH-LEVEL SECURITY REQUIREMENTS FROM SECURITY PROPERTIES.

| Security Property | OWASP High-level Security Requirement |
|---|---|
| Confidentiality (C) | C1. Data shall be protected from unauthorized observation and disclosure both in transit and when stored. |
| | C2. System sessions shall be unique to each individual and cannot be shared. |
| | C3. System sessions are invalidated when timed out during periods of inactivity. |
| | C4. TLS protocol shall be used where sensitive data is transmitted. |
| | C5. System shall use strong encryption algorithm at all times. |
| Integrity (I) | I1. Any unauthorized modification of data must yield an auditable security-related event. |
| | I2. All input is validated to be correct and fit for the intended purpose. |
| | I3. Data from an external entity shall always be validated. |
| Availability (A) | A1. The application server shall be suitably hardened from a default configuration. |
| | A2. HTTP responses contain a safe character set in the content type header. |
| | A3. Backups must be implemented and recovery strategies must be considered. |
| Identification & Authentication (IA) | IA1. Users are associated with a well-defined set of roles and privileges. |
| | IA2. The digital identity of the sender of a communication must be verified. |
| | IA3. Only those authorized are able to authenticate and credentials are transported and stored in a secure manner. |
| | IA4. Passwords treatment must include complex passphrases, options to recover and reset the password and default passwords not allowed. |

For each user story, the generated reading technique focuses the reviewer to verify whether its security specifications contain any of the defect types. This happens when reviewers check the security specifications against the OWASP high-level security requirements of the linked properties. To reach this, the reading technique contains a set of verification questions to help identifying the different defect types. Table 6 shows the questions.

TABLE 6. VERIFICATION QUESTIONS FOR THE DIFFERENT DEFECT TYPES.

| Type of defect | Question |
|---|---|
| Omission | When comparing the security specifications with the OWASP high-level security requirements, are there high-level security requirements or characteristics that were not specified? |
| Ambiguity | Does any security specification allow for multiple interpretations? |
| Inconsistency | Are there two or more security specifications in conflict with one another? |
| Incorrect Fact | Is there any security specification stating information that is not true under the conditions specified for the system? |

*Phase 2: Following the Reading Technique for Detecting Defects.* Typically, security defects in early phases result in security vulnerabilities in later development phases. Hence, providing complete, consistent, unambiguous, and correct security specifications throughout the life cycle improves the chances for a higher quality software system. The aim of this phase is to guiding the reviewer on finding such requirements defects. Using the generated reading technique, reviewers can follow the instructions and answer the questions in order to look for defects.

To facilitate the review task, our approach rewrites the OWASP high-level security requirements in a way so that inspectors can easily identify certain security aspects. For instance, we use the AND logical connector in capital letters to get the attention of the reader and indicate that both aspects must be considered to satisfy the high-level security requirement. *E.g., "C1. Data shall be protected from unauthorized observation or disclosure both in transit AND when stored"*. In this case, if the specifications were well specified, they must be considering security aspects related to data protection both in transit and in storage. Otherwise, there is an omission defect.

Furthermore, the security statements also present examples for some concepts in order to give inspectors an idea about the context of the OWASP high-level security requirement. An example follows: *"I2. All input (e.g., query parameters, string variables, REST calls and cookies) is validated to be correct and fit for the intended purpose"*.

That way, reviewers are provided with a focused reading technique that should increase their performance during the review. We give a motivational example on how the reading technique is used in the next section.

### D. Motivational Example

In the following, we demonstrate the application of our approach in an exemplary setting. Table 7 shows a user story and its set of security specifications with some defects commonly applicable to any agile software project.

TABLE 7. INPUT OF THE APPROACH AS AGILE REQUIREMENTS SPECIFICATIONS.

| User Story | Security Specification |
|---|---|
| 1. As a customer, I want to be able to export my personal information so that I can use it in other systems. | 1. The system shall ensure that there is no residual data exposed. |
| | 2. The system shall store credentials securely using the AES encryption algorithm. |
| | 3. The system shall use the RSA encryption algorithm to protect all data all the time. |
| | 4. The system shall inactivate a session when it exceeds certain periods of inactivity. |
| | 5. The system shall encrypt the roles and privileges of the system. |

With the user story in sight, the framework extracts the relevant words and matches the related security properties. In this case, the extracted words are "export" and "system". Thereafter, the framework can verify whether some security property is related to the extracted words. In this case, according to Table 4, "export" matches confidentiality, while "system" does not match any of the security properties. Therefore, our approach can propose OWASP high-level security requirements for the confidentiality security property to be verified for the user story (*cf.* Table 5, confidentiality).

Our approach then generates the defect reporting form by showing the user story with its security properties and its

OWASP high-level security requirements, and the verification questions. Thus, inspectors know which security aspects they should verify. The verification process starts at this point. By having inspectors responding to the verification questions looking for defects, we expect to obtain valuable insights from them on the quality of the security specifications. An exemplary enactment of answering these questions follows.

*When comparing the security specifications with the OWASP high-level security requirements, are there high-level security requirements or characteristics that were not specified?* In this case, 3 out of 5 OWASP high-level security requirements linked in Table 5 are related or make sense to the security specifications. This means that two OWASP high-level security requirements (C2 and C4) are not covered in the security specifications. Therefore, we have detected two defects that should be marked as "omission".

*Does any security specification allow for multiple interpretations?* The answer is affirmative. Security specification 4 (SS4) reflects a weak statement as the amount of time with respect to "certain periods of inactivity". It could be hours or seconds. Thus, we have identified a defect related to ambiguity.

*Are there two or more security specifications in conflict with one another?* In this case, the answer is affirmative. Security specification 2 (SS2) and security specification 3 (SS3) are in conflict because SS3 indicates to encrypt all data using the RSA algorithm. Nevertheless, SS2 indicates to protect credentials, which are also data, using the AES algorithm. Thus, we have identified a defect related to inconsistency.

*Is there any security specification stating a characteristic that cannot be true under the conditions specified for the system?* Security specification 5 (SS5) is not correct because concepts/functionalities of the system cannot be encrypted. The concept "encrypt" is not correct in the statement.

Finally, the reviewers fill out the defect reporting form that summarizes the identified defects. Table 8 presents the output of the verification using the reading technique (a similar empty Table is generated and provided to the inspector during the task). Note that OM column is related to OWASP high-level security requirements that were omitted. The other columns are related to the remaining defect types listed in Table 2.

TABLE 8. DEFECT REPORTING FORM OF THE APPROACH.

| User story | Security Property | OWASP High-level Security Requirement | OM | AM | IS | IF |
|---|---|---|---|---|---|---|
| US1 | Confidentiality | C1. Data shall be protected from unauthorized observation AND disclosure both in transit AND when stored. | | | | |
| | | C2. System sessions shall be unique to each individual AND cannot be shared. | X | | | |
| | | C3. System sessions are invalidated when timed out during periods of inactivity. | | SS4 | SS2 SS3 | SS5 |
| | | C4. TLS protocol shall be used where sensitive data is transmitted. | X | | | |
| | | C5. System shall use strong encryption algorithm (e.g., DES, AES, RSA) at all times. | | | | |

In summary, this table indicates that the security specifications related to the user story (US1) contain six defects. Two out of them were marked as omission (OM) because none of the OWASP high-level security requirements (C2, C4) is related to the security specifications. The rest of the defects (3) are related to ambiguous, inconsistent and incorrect fact defects. In this case, security specification 4 (SS4) was marked as ambiguous, security specification 2 and 3 (SS2, SS3) were marked as inconsistent and security specification 5 (SS5) was marked as incorrect.

## IV. EXPERIMENT

For a better understanding of the feasibility and impact of using our approach for reviewing security-related aspects in agile requirements specifications of web applications, we conducted a controlled experiment in academic settings with students. The choice of a controlled (difference-making) experiment in a laboratory setting is intended rather than being opportunistic, because we are specifically interested in investigating specific phenomena in isolation as a preparation for scaling our implementation (and evaluation) up to practice. To this end, we followed the guidelines proposed by Jedlitschka and Pfahl [27] and Wohlin et al. [44] for reporting the experiment.

### A. Experiment Goal

We defined our study goal following the GQM (Goal Question Metric) template [4]. The goal is shown in Table 9.

TABLE 9. EXPERIMENT GOAL BASED ON GQM.

| Analyze | the reading technique generated by our approach |
|---|---|
| for the purpose of | Characterization |
| with respect to | the effectiveness, efficiency, usefulness and ease of use of the approach |
| from the point of view | researcher (on the measured effectiveness and efficiency) and inspectors (on the perceived usefulness and ease of use) |
| in the context of | novice inspectors using the technique on agile requirements specifications, when compared to directly using the OWASP high-level security requirements and the requirements defect types. |

Based on our goal, we formulated two Research Questions (RQ). *RQ1. Does the approach have an effect on defect detection effectiveness and efficiency?* and *RQ2. How do the inspectors perceive the usefulness and ease of use of the approach?* The variables used to answer these RQs are described in detail in Subsection C.

### B. Experiment Context

The experiment was conducted in two trials, involving students enrolled in Software Engineering classes at (*name omitted for double-blind conformance*). It is noteworthy that we also carried out a pilot study with two independent volunteers. The aim was to evaluate the overall (particularly technical) feasibility, time, adverse events, and improve the experiment materials before the experiment trials.

We created the specifications based on typical customer requests for developing web applications, e.g., sending sensitive information to other software systems and deleting data. When doing so, we relied as an orientation on SR specifications from real industrial software projects as used by our industry partners. Our goal is to increase the similarity to the studied population units, but did not use real specifications verbatim in our setting for confidentiality reasons.

Our SR specification contains a set of user stories in this format: As a [Role], I want [Feature], so that [Reason]. The document also contained the related security specifications with seeded defects. The requirements were peer-reviewed by three independent researchers before conducting the experiment trials.

To avoid the defect seeding to represent a confounding factor, the type and amount of seeded defects to evaluate the suitability of our approach was carefully considered. Table 10 shows the distribution of the seeded defects per user story. In total, 13 defects were seeded. The representativeness of the defects was peer reviewed by three independent researchers.

TABLE 10. DISTRIBUTION OF THE SEEDED DEFECTS.

| User Story | Omission | Ambiguity | Inconsistency | Incorrect Fact | Total |
|---|---|---|---|---|---|
| US1 | 2 | 2 | 2 | 1 | 7 |
| US2 | 2 | 2 | 1 | 1 | 6 |

### C. Variables Selection

The independent variable in the experiment is the treatment applied by the groups in order to find defects in the SR specifications. While the control group received OWASP high-level security requirements and a list of defect types to be found, the experimental group received our proposed reading technique.

Regarding dependent variables, we used effectiveness and efficiency, defined as follows. *Effectiveness* is expressed as the ratio between the number of real defects found and the total of seeded defects in the documents. On the other hand, *Efficiency* refers to the ratio between the number of real defects found and the time spent in finding them. For these variables, we collected quantitative data to test the hypotheses presented in Subsection D. We also collected qualitative feedback in a follow-up questionnaire built on basis of the *Technology Acceptance Model* (TAM) questionnaire [12]. The aim was to gain insights about the perceived usefulness and ease of use of the approach. TAM has been extensively used and validated for this purpose [40].

### D. Hypotheses

Using the variables described in the previous subsections, we defined the following hypotheses.

- **Null hypothesis on effectiveness (H0a):** There is no difference in terms of effectiveness when using both techniques.
- **Alternative hypothesis on effectiveness (H1a):** There is a difference in terms of effectiveness when using both techniques.
- **Null hypothesis on efficiency (H0b):** There is no difference in terms of efficiency when using both techniques.
- **Alternative hypothesis on efficiency (H1b):** There is a difference in terms of efficiency when using both techniques.

### E. Selection of Subjects

Our subjects were intended to represent novice inspectors. We selected subjects, by convenience, from classes on Software Engineering at *(name omitted for double-blind conformance)*, involving 25 undergraduate (first trial) and 8 graduate students (second trial).

We characterized the subjects by their experience and knowledge on four areas: agile software development, requirements engineering, software security and inspections. As a result, we found that the majority of students had low level of security experience and knowledge. We also found that participants are not familiar with requirements inspection techniques. Hence, they match our intended profile. Details of the characterization are presented in the online material [4]. Aiming at mitigating threats to validity concerning the distribution of subjects between groups, we used the characterization and applied the principles of balancing, blocking and random assignment (within the blocks) [44]. Hence, students who demonstrated knowledge on software security (4 out of 33) were separated and distributed equally into the control and the experimental groups of each trial.

Subjects who found less than 10% of the defects were discarded as outliers, because, in our understanding, their results reflect misunderstanding or lack of interest in the review activity. Following this idea, we had to discard 5 participants from the first trial. In the second trial, it was not necessary to discard participants.

### F. Experiment Design

Our experiment is composed of one factor with two treatments: (1) using our proposed reading technique and (2) directly using the OWASP high-level security requirements and a list of defect types to be found. The study design is composed by a set of five activities distributed into three phases. Figure 2 shows how all the phases of the experiment were organized.

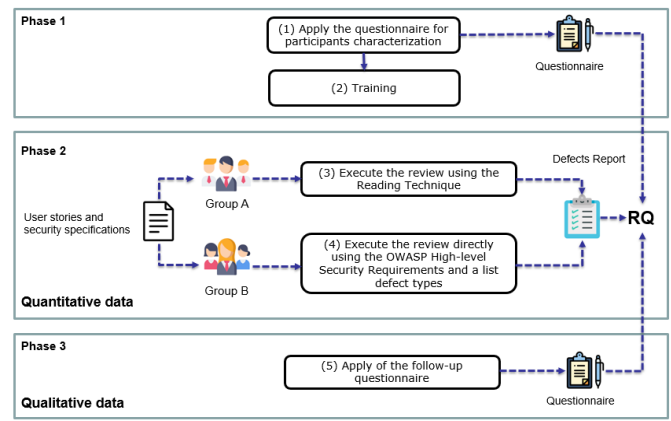

Fig. 2 Experimental Design

In the first phase, all the students filled out a characterization questionnaire with questions about their expertise in the topics related to the study. They also received training to introduce the main topics. In the second phase, we obtained quantitative data by conducting two trials. The students of each trial were divided into two groups in order to evaluate the performance by executing the review using or not our approach. Finally, in the third phase, the participants of the experiment gave us feedback on the execution of the experiment. The instruments used to conduct the experiment are further described in the following.

---
[4] https://doi.org/10.5281/zenodo.2602205

### G. Instrumentation

In order to make available the experiment package with all the instruments and the results, we followed open science policies that are available at Zenodo[4]. The instruments used in the experiment are presented as follow.

*Characterization questionnaire*. The overall goal of the questionnaire is to characterize the experience of the students. The answers were obtained through the questionnaire with the aim at identifying some key characteristics about four knowledge areas: agile software development, requirements engineering, software security and inspections.

*Follow-up questionnaire*. This questionnaire was based on TAM [40]. We wanted to know if the approach was useful for them and if they found it easy to use. Additionally, we included some open text questions to gather participant feedback about difficulties, benefits and disadvantages of using our approach.

*Training*. The training material focused on the OWASP security properties and high-level security requirements and on the defect types. No specific training was provided on using the reading technique; therefore, the feasibility of using it without specific training was also indirectly evaluated.

*Task description*. This description explains the participants about the received material and asks them to conduct the review according to their treatment, filling out the defect reporting form. Both treatments received the same requirements specification. For one treatment, the reading technique was generated according to the user story description and its related OWASP high-level security requirements. For the other treatment the list of OWASP security properties and their related high-level security requirements was provided together with the description of the defect types to be located.

*Defect reporting form*. This form was used by participants to record the start and end time of the inspection, as well as the defects by location, type and description. The defect reporting form for the experimental group was the one generated for applying the reading technique, similar to the one shown in Table 8.

### H. Experiment Operation

The experiment was executed along two days. On the first day, the subjects answered the characterization form in order to allow dividing them into experiment groups. On the second day prior to the execution of the experiment, concepts of the OWASP security properties, high-level security requirements, and relevant types of defects were reviewed by participants in a training session. After that, the inspection was conducted as follows.

All subjects had up to 60 minutes to carry out the review. During the experiment, the control group used the OWASP high-level security requirements and a list of defect types as support during the inspection. The experimental group used our proposed reading technique. When the subjects from both groups finished the task, they had to fill out the follow-up questionnaire. The first and second trial were conducted with the undergraduate and graduate students, respectively. The aim was to obtain valuable feedback from the first trial to consider improvements in the second.

### I. Threats to Validity

We organize the information on the threats to validity as suggested by [44].

Internal validity. First, aiming to avoid personal bias, we used researcher triangulation and all collected data was analyzed by three researchers. Second, we characterized all the participants of the experiment. The characterization allowed us to apply the blocking principle, which consisted of removing characterization-related confounding factors by distributing the participants so that these characteristics were equally distributed among the groups. Random assignment was employed for participants with similar characteristics. Balancing was applied to assure that group sizes were equally distributed. Finally, participants received the exact same training.

*Construct validity*. For our quantitative analysis, we used metrics (effectiveness and efficiency) that are commonly used in inspection experiments. The qualitative analysis, on the other hand, relies on the TAM, which has also been widely used and evaluated [40].

*Conclusion (statistical) validity*. This type of validity is strongly related with the sample size and the statistical methods employed during the data analysis. Besides the analysis procedures previously described, the statistical hypothesis testing methods were chosen according to the sample distribution. Additionally, besides statistical significance, we also measured the treatment effect sizes.

*External validity*. As we planned to conduct a limited amount of trials with a limited amount of subjects, the experiment package is available for external replications. Regarding subject representativeness, we used students to represent novice inspectors. Using students as participants remains a valid simplification of real-life settings needed in laboratory contexts [17]. Regarding the used objects, we also peer-reviewed the requirements specifications and the seeded defects in terms of their representativeness.

## V. RESULTS

In the following, we present the results of our controlled experiment. In addition, we describe the data collection and analysis procedure to get the answer to the research questions.

We executed three steps to collect the data necessary for answering our research questions. First, we collected the number of real defects found. Based on this data it is possible to evaluate the performance of the treatments in terms of effectiveness. This metric is used to partially answer RQ1. Furthermore, we collected the time spent for detecting defects. This allows us to compare the performance of the treatments in terms of efficiency, which is the metric that complements the answer to RQ1. Finally, we collected answers from the follow-up questionnaire. This questionnaire allowed us to receive feedback on the perceived usefulness and ease of use. This provides information for answering RQ2.

Regarding analysis procedure, we conducted both, quantitative (RQ1) and qualitative analyses (RQ2). Related to quantitative analysis, descriptive statistics of our experiment trials were based on the above described metrics. Thereafter, to answer RQ1, statistical hypothesis testing was applied. Our analysis was conducted using the statistical tool RStudio version 1.1.423. For hypotheses testing, the data analyst used the Mann-Whitney non-parametrical test with $\alpha = 0.05$. This choice of statistical significance level and test was motivated by the small and independent samples. For the qualitative analysis, on the other hand, we present the frequencies related to the TAM questionnaire. Additionally, grounded theory coding activities were applied to obtain further understand difficulties, benefits, and disadvantages.

**RQ1: Does the approach have an effect on defect detection effectiveness and efficiency?**

We wanted to understand the potential of the approach to detect security defects in agile specifications of web applications. For this, we compared the performance of using our approach versus the review based on the OWASP high-level security requirements and a list of defect types. Regarding the effectiveness, Figure 3 shows the distribution of the amount of defects found by the students in each trial.

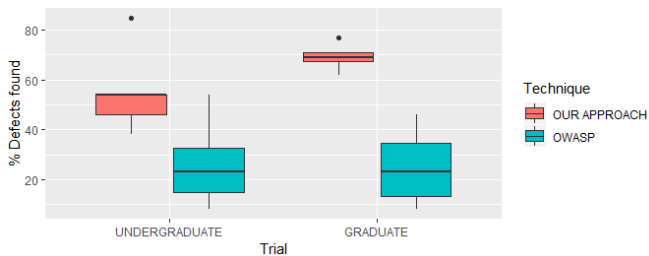

Fig. 3 Defect detection effectiveness

It is possible to observe that in both trials our approach was more effective than the other review. In the first trial, the students who used our approach identified, in median, 54% of the defects while those who did not use it identified 23%. The difference was even stronger when observing the performance of the graduate students (second trial). Graduate students who used our approach identified, in median, 69% of the defects versus 23% identified by the students who did not use it. The improvement for the graduate trial may have happened because we slightly modified the defect reporting form for the experimental group in the second trial to ease understanding and fulfillment (equivalent to the one shown in Table 8). The reason was that several inspectors mentioned in the first trial's follow-up questionnaire that the defect reporting form was confusing. In other words, we improve the design of the defect reporting form, while it remained capturing the exact same information.

We also wanted to *test our null hypothesis on the effectiveness (H0a)*, i.e., we checked whether these differences were significant. The results of the tests allowed to reject the null hypothesis for both trials (p-values of 0.002 and 0.012 for the first and second trial, respectively); this means there is significant difference in terms of effectiveness between our approach and the other defect-based technique. In addition, we calculated the Cohen's effect size [41] for both trials (3.46 and 2.24 for graduate and undergraduate students, respectively). Thus, we can partially answer RQ1: Our approach has a positive impact on defect detection effectiveness with very large effect size.

After knowing the effectiveness of our approach, we can question its efficiency by analyzing the defects found per hour by the inspectors. Figure 4 shows the distribution of the efficiency of the students involved in the experiment per trial.

Note that the efficiency follows the same pattern of the effectiveness. That is, the amount of defects found per hour by the inspectors who used our approach was greater than the one of the inspectors who used the other technique. In the first trial, the median of our approach efficiency was 15 defects found per hour (we seeded only 13 defects, but participants took less than one hour to complete their tasks), while the median of the other one was four. In the second trial, the median of our approach efficiency increased to 21 defects found per hour versus four defects found per hour by the control group. *I.e.*, inspectors who used our approach identified defects faster.

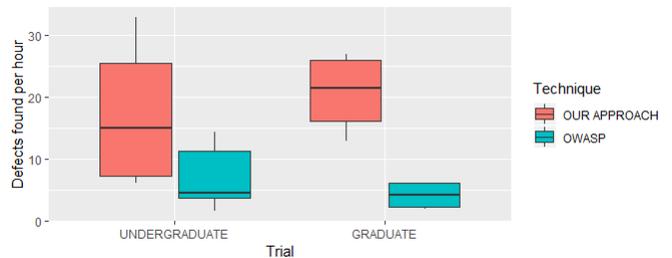

Fig. 4 Defect detection efficiency

Regarding the statistical hypothesis testing for efficiency, we found that the Mann-Whitney Test suggests rejecting our second null hypothesis (p-values of 0.02 and 0.01 for the first and second trial respectively). This means there is significant difference in terms of efficiency between our approach and the OWASP review. Additionally, the relevance of this difference (Cohen's effect size [41]) was very large for both trials (1.56 and 3.29 for undergraduate students and graduate students respectively). With this information, we can fully answer RQ1. Our approach has a positive impact on security defect detection effectiveness and efficiency, when compared to directly using the OWASP high-level security requirements and the requirements defect types as basis for verification.

**RQ2: How do the inspectors perceive the usefulness and ease of use of the approach?**

After inspectors reviewed the security specifications, we asked if they found the approach useful and easy to use. Through the TAM questionnaire, we wanted to know about their perceptions on using the approach. All those who used our approach strongly agreed (75%) or partially agreed (25%) that their performance improved in some way (find defects faster). This last one perception may be strongly related with the lack of experience and security knowledge of the inspectors and difficulties faced by them when conducting the review. For instance, one inspector stated the following: "The review may be exhausting and time consuming due to the task description document is not lightweight". The inspectors also mentioned some difficulties faced such as "confusing defect reporting form" and "requirement specifications are too abstract".

Regarding ease of use, again all those who used our approach strongly agreed (17%) or partially agreed (83%). The large number of partial agreements indicates that improvements could be introduced in order to facilitate the use of the approach. According to the follow-up questionnaire, the inspectors proposed some points that may improve the understanding of the approach such as providing a lighter document, modifying the design of the defect reporting form and showing an example of how to fill it out correctly. This is convincing because some inspectors reported difficulties to adopt the approach. On the other hand, three undergraduate students mentioned that once the review process using the approach is understood, the detection of defects is simple.

VI. DISCUSSION

In this section, we discuss potential practical implications of our research for practitioners and researchers. For practitioners, our approach provides a way to detect defects related to security aspects that should be specified. According to the results of the controlled experiment, we are confident to say that in principle our approach supports novice inspectors by providing guidance in applying a reading technique that will help them to identify defects in agile requirements specifications of web applications. In addition, we designed the approach in such a way that it works without expensive review cycles, aligned with the agile philosophy. We see three main potential benefits of this approach. First, narrowing the security knowledge gap that exists between experts and novice inspectors. Second, the reading technique provides a strong focus on security aspects. In this way, the team can avoid discussing obvious issues and focus on important, difficult, security-specific aspects of the review. Third, we saw that our approach provided positive results regarding the performance of individual inspectors when conducting the review.

For researchers, this work contributes already in closing an important literature gap that exists with respect to security requirements verification in the agile context. Based on the scarce literature on the topic, we believe that our approach constitutes an interesting starting point to discuss in depth verification activities centered on security in the agile context. For us, the results strengthen our confidence in further extending our approach to scale its usability up to practical settings covering a full, too-supported process integration, which was not (and could not be) in scope of a development in our research-centric environment.

VII. LIMITATIONS

We concentrated on a set of concrete security properties and high-level security requirements from the OWASP (matching security sub characteristics also described in the SQuaRE quality model [26]). There are several security standards and guidelines that are different to the ones provided by the OWASP project [33]. This means that we could complement the security vision of our approach with other standards.

Moreover, given the complexity of working with natural language in RE, there is a limitation related to the completeness of the keyword repository needed to link the user stories with the security properties. To deal with this, we decided to consider as many synonyms as possible regarding the initial set of keywords.

We are also aware that our security specifications constitute a limitation of the study. In a perfect scenario, we would have security concerns specified by companies or independent practitioners, but often this information is restricted. Therefore, we invested our best efforts to carefully create and verify the specifications on their representativeness. Nevertheless, external replications, including a wider range of user stories and security specifications, are needed to improve external validity of our results.

VIII. CONCLUDING REMARKS

We proposed an approach for reviewing security related aspects in agile requirements specifications of web applications. The approach considers user stories and security specifications as inputs and involves applying NLP in order to relate those user stories to candidate security properties and high-level security requirements proposed by the OWASP. As a result, the approach provides a user story focused reading technique that can be used to support the manual inspection of agile specifications.

We validated the approach by designing and conducting two trials of a controlled experiment. The purpose was to validate the feasibility of using our approach for detecting defects related to security aspects in web applications. In these trials, our approach had a positive effect on defect detection effectiveness and efficiency. Inspectors who used our approach identified more defects in less time than inspectors who conducted the inspection using the OWASP high-level security requirements and a list of defect types. In principle, participants found the reading technique generated by our approach useful and easy to use.

Future work includes to evaluate and understand the performance of using our approach in industry settings and when compared to other inspection techniques (e.g., PBR-security [10], even though it was not designed for the agile context). Furthermore, we consider it important to follow an exemplary process integration, e.g., by further automatizing the approach within the FESCAR framework, in such a way that applying the reading technique could be guided by the framework. This could help mitigating difficulties mentioned by the participants of the experiment. In addition, we are aware that our repository of keywords is still limited and that our approach could be evolved beyond exact keyword matching.